\documentclass[twocolumn,showpacs,aps,prl,groupedaddress,amssymb,amsmath,nobalancelastpage]{revtex4}

\usepackage{graphicx}% Include figure files
\usepackage{longtable}% Include figure files

\begin{document}

\title{Viscous shear-banding in foam}

\author{Kapilanjan Krishan}
\author{Michael Dennin}
\affiliation{Department of Physics and Astronomy, University of
California at Irvine, Irvine, California 92697-4575}

\date{\today}

\begin{abstract}

Shear banding is an important feature of flow in complex fluids.
Essentially, shear bands refer to the coexistence of flowing and
non-flowing regions in driven material. Understanding the possible
sources of shear banding has important implications for a wide
range of flow applications. In this regard, quasi-two dimensional
flow offers a unique opportunity to study competing factors that
result in shear bands. One proposal for interpretation and
analysis is the competition between intrinsic dissipation and an
external source of dissipation. In this paper, we report on the
experimental observation of the transition between different
classes of shear-bands that have been predicted to exist in
cylindrical geometry as the result of this competition [R. J.
Clancy, E. Janiaud, D. Weaire, and S. Hutzlet, Eur. J. Phys. E,
{\bf 21}, 123 (2006)].
\end{abstract}

 \maketitle

%\noindent Some papers of relevance:\\
%Dublin PRL 2006: \cite{jwh06}\\
%Denkov PRL 2008: \cite{dtgal08}\\
%Dublin EPJE 2006: \cite{cjwh06}\\
%Chris,Scott PRE 2006: \cite{csd06}\\
%Yuhong PRE 2006: \cite{wkd06a}\\

It has long been assumed that flowing and non-flowing regions can
exist in driven complex fluids, especially in situations in which
the stress is inhomogeneous. This behavior is often discussed in
the context of {\it yield stress} materials \cite{BOOKS,MMB06}.
These are materials for which a critical stress is required for
flow to be generated. Naturally, in a situation in which the
stress varies from values below to above the critical, or yield,
stress, one will have coexistence of flow and no flow. More
recently, experimental studies have revealed a rich variety of
shear-banding situations that do not conform to the
``traditional'' yield stress picture of shear bands
\cite{HBV99,MDKENJ00,LBLG00,CRBMGH02,VBBB03,KD03,LCD04,BMC06,gsd06,CDSBW06,jwh06,cjwh06,KD07a,KMH08}.

One of the challenges to an increased understanding of shear
banding in complex fluids is the large number of parameters and
observed behavior. There is clearly no {\it single} cause of shear
banding, and one must sort out the mechanisms for a range of
different {\it qualitative} behaviors. Among the interesting range
of behavior and models, we are going to highlight a few cases
because of their relevance to the work discussed here and because
they highlight the importance of distinguishing between
qualitatively different types of shear banding. It is important to
realize that these cases are not necessarily mutually exclusive.
We will first consider issues related to the particulate nature of
foams.

When considering shear banding in many complex fluids, one has to
determine if a continuum model will be relevant or if one has to
consider in detail the fact that the fluid consists of macroscopic
``particles''. For example, foam is gas bubbles in the micron to
millimeter range surrounded by liquid walls. Granular matter is
similar size particles with gas. Experimentally, a transition to a
``discrete'' flow regime has been observed for extremely narrow
shear bands on the order of 10 - 20 particles \cite{RBC05,gsd06}.
In this case, an interesting feature of the shear band is that the
width becomes independent of the external driving velocity. This
is fundamentally different behavior from continuum shear bands,
where the width scales with the driving.

Another qualitative signature of different types of shear
localization is the continuity, or lack there of, of the {\it rate
of strain} at the transition from flow to no-flow. Continuous
versus discontinuous transitions represent potentially different
physics. This is supported by observations of a discontinuous
transition in the rate of strain in a range of experimental
systems \cite{D08}, including both three-dimensional systems and
in quasi-two dimensional systems (bubble rafts)
\cite{LCD04,gsd06}. For simple cases, the origin of the
discontinuity may be understood in terms of a competition between
clustering of particles and the breakdown of cluster by shear
\cite{MRMB08}, but a complete theoretical understanding of
mechanism for selecting between continuous and discontinuous
transitions is still lacking.

Another qualitative feature of shear bands, especially in foams,
is the fact that the velocity localization is often correlated
with the localization of other dynamical features in the system.
For example, flow in foams is dominated by local rearrangements of
the bubbles (T1 events). The degree of correlation between
velocity and T1 localization, and the possible role of T1 events
as a source of shear localization, are an important feature of
shear bands. Models have suggested that shear bands can result
when local stress inhomogeneities are long lived and produce
spatio-temporal correlations in the occurrence of plastic events
in the flow (often associated with T1 events) \cite{KSD07b}. This
tends to result only in continuous shear bands, so it serves as an
alternative model to ones that describe discontinuous transitions.
More recently, a useful direction for exploring the role of T1
events in focusing the flow has been proposed using surface
evolver \cite{CW08}. This work has shed interesting light on the
role of polydispersity in controlling the degree of localization.

As one starts to consider the various causes of shear
localization, studies with quasi-two dimensional systems present
an interesting option: role of external versus internal
dissipation. This case has been of interest for a class of systems
involving a single layer of bubbles. For these systems, there are
three common configurations for the boundaries that are used to
confine the bubbles to a plane. There is the classic bubble raft
consisting of a single layer of bubbles at the air-water interface
\cite{BL49}. This has recently been labelled the liquid-air (LA)
geometry \cite{VC05}, which is a useful distinction when making
comparison with other common quasi-two dimensional geometries for
foams. One can also confine the bubbles between glass plates in a
Hele-Shaw geometry, referred to as glass-glass (GG). Finally, one
can study a mixed system using bubble on a water surface with a
top glass plate, the liquid-glass geometry (LG). Detail studies of
the dynamics of individual bubbles in these geometries have helped
our understanding of the differences between the systems
\cite{VC05}. Experimental studies in all three systems have
established conclusively that dissipation between the bubbles and
the boundaries can produce shear bands~\cite{wkd06a,KD03,KMH08}.
Various theoretical models and simulations have pointed to the
importance of the details of the competition between bubble-bubble
dissipation and bubble-boundary dissipation
\cite{wkd06a,jwh06,cjwh06,KMH08}. A particularly interesting
prediction exists for the case of the circular Couette geometry
\cite{cjwh06}.

Circular Couette flow occurs for a material confined between two
concentric cylinders. One can drive flow from either the outer or
inner boundary. An important characteristic of Couette flow (in
the absence of any external dissipation) is the fact that the
stress decays as $1/r^2$ as one moves from the inner to the outer
cylinder, independent of the driving. This establishes a natural
situation for ``classic'' shear banding of a yield stress fluid,
with flow occurring near the inner cylinder where the stress is
the largest. An interesting feature of studies with bubble rafts
is that this classic case is generally not observed. Instead, the
shear bands are either discontinuous in the rate of strain or
examples of discrete flow~\cite{LCD04,gsd06}. In this paper, we
consider the case where external dissipation does exist, which
modifies the spatial dependence of the internal stress, and
explore the qualitative features of the shear banding.

Recently, theoretical studies of the impact of dissipation from an
external boundary have predicted a range of striking behavior in
Couette flow driven by the outer cylinder \cite{cjwh06}. For
sufficiently high external dissipation, one should observe a
series of transitions from a single shear band at the inner
cylinder to shear bands at the inner and outer cylinder, and
ultimately for high enough velocities of the outer cylinder, a
shear band exclusively at the outer cylinder. This effect is
possible because of the velocity dependent dissipation of the
confining boundaries. In this paper, we report on an experimental
confirmation of the transition from single shear bands to two
shear bands in the flow of a bubble raft.

\begin{figure}
\includegraphics[width=6cm]{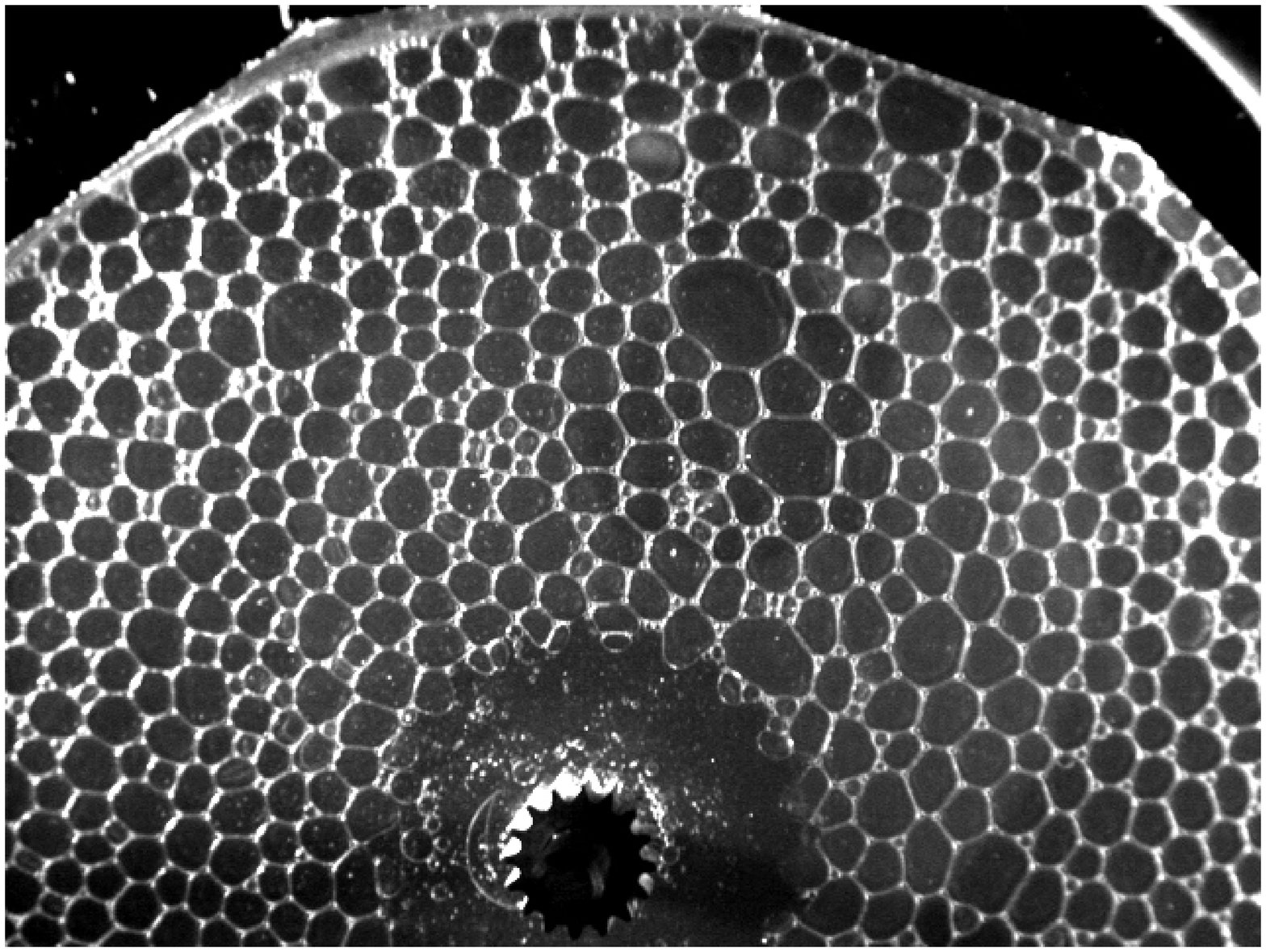}\\
\includegraphics[width=6cm]{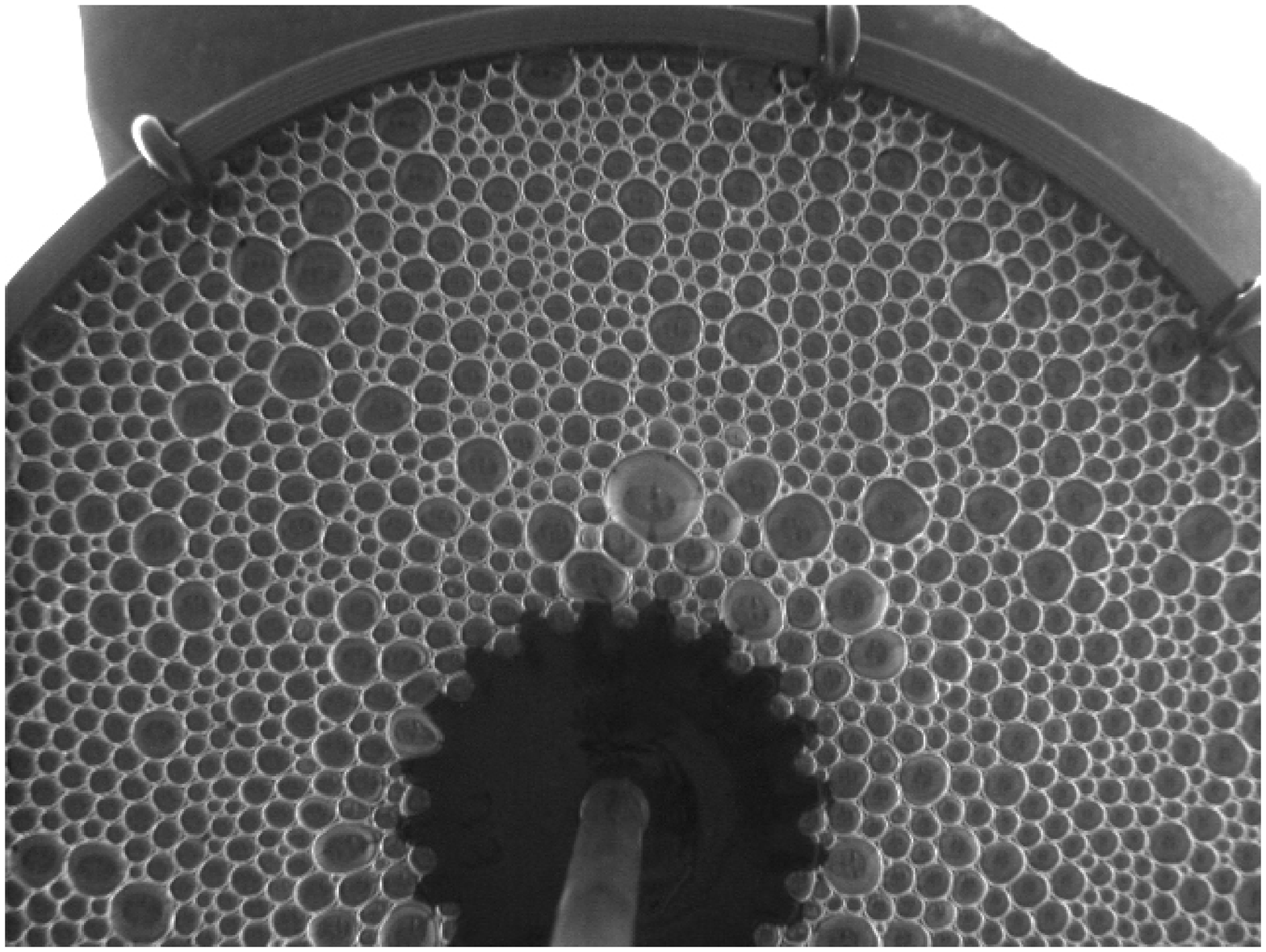}\\
\includegraphics[width=6cm]{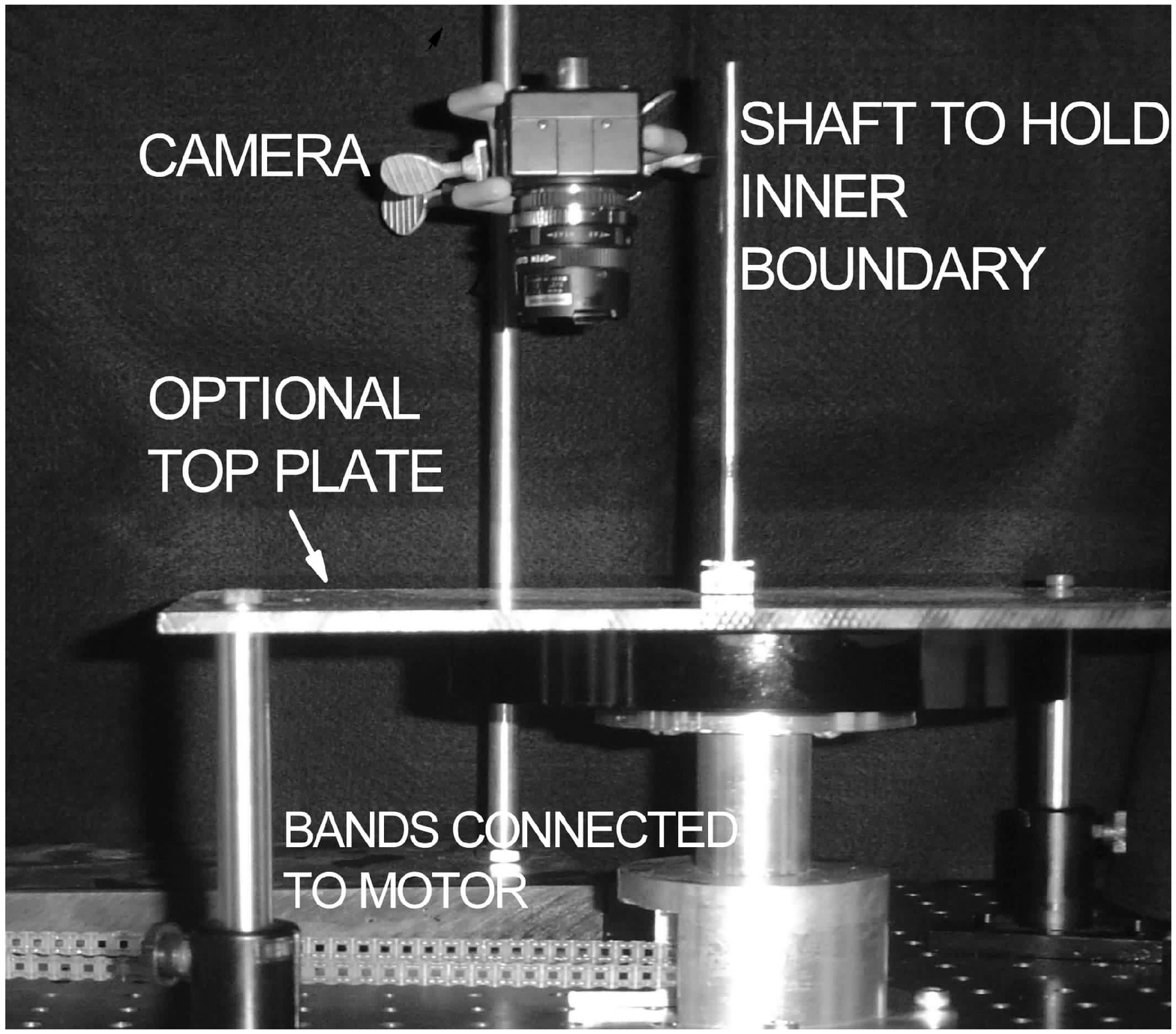}
\caption{The top two panels are images of the bubble raft with
(top panel) and without (middle panel) a top. The distance from
the center of the trough to the outer wall is $16\ {\rm cm}$. The
lower panel is a photograph of the apparatus illustrating the
experimental geometry.} \label{apparatus}
\end{figure}

Our experimental setup utilizes bubble rafts in a Couette
geometry. We are able to run both with and without a top boundary
for the bubbles. For the bubble raft, we find that the flow is
driven predominately from the rotating outer boundary, and the
drag due to viscous dissipation between the bubbles and the
aqueous subphase is relatively negligible. For our Couette system,
there is a circular dish ($160\ {\rm mm}$ diameter, $20\ {\rm mm}$
deep) mounted from below on a shaft. A corrugated disk,
approximately $45\ {\rm mm}$ in diameter was set concentrically
within the circular dish. The shaft was supported by a set of
bearings and connected using gears and driving bands to a variable
speed stepper motor. The circular dish was filled with a
homogeneous solution of high purity water, glycerol and miracle
bubble solution in a 80:15:5 volume ratio. A stream of nitrogen
gas was bubbled through the solution using a needle to generate a
bubble raft in the spacing between the walls of the circular dish
and inner disk. The bubbles in this 2D foam are polydisperse, with
a diameters ranging between 2~mm to 9~mm. The circular dish was
rotated at different angular velocities in order to generate
differing amounts of shear within the 2D foam. The bubble raft was
imaged from above using a standard CCD camera and digital images
were saved to a computer every $0.1\ {\rm s}$ for a total of 4000
images. A standard PIV technique was used to track the bubbles and
compute the azimuthal velocity profile as a function of radial
position. The velocity profiles were found to have converged over
the observation period for all the angular velocities studied. An
image of the apparatus is given in Fig.~\ref{apparatus}.

The experiment was run in two configurations: a) with the bubble
raft open to the atmosphere and b) with a 1/4 inch thick,
transparent polycarbonate sheet in contact with the bubbles on
top. With the addition of the top plate, a relatively uniform
viscous drag is generated between the bubbles and the top plate.
This drag force is velocity dependent. In both cases, the bubbles
at the inner barrier are held stationary at the corrugated
barrier. In case (a), slip at the outer barrier is eliminated by
the insertion of 2~mm thick protrusions along the outer wall. In
case (b), the viscous drag between the top plate and the foam was
stronger than that with the outer wall, and we were unable to
eliminate slip between the outer wall and the last layer of foam.
In this case, a corrugated barrier was found to be insufficient to
hold the bubbles in place, instead causing artifacts by frothing
of the bubble raft even at low angular speeds. Therefore, in case
(b), the outer wall was moved at a higher angular velocity than
the bubbles in contact with it. The viscous drag between the
bubbles and the liquid subphase is not a significant contribution
to the dynamics of the flow in comparison to that of the top plate
and the outer wall. This is verified by checking that the bubbles
do not undergo any rotational motion when only in contact with the
inner barrier and the rotating subphase (without touching the
outer wall, or the upper plate). A limitation in our experiment is
that at high angular velocities with a top plate, a frothy
solution was formed at the outer barrier that changed the
dispersity of the foam from its initial condition. In the data
presented in this paper, we limit our angular velocities to a
regime where the froth formation does not occur significantly. The
qualitative difference in the flow profile was not seen to be
strongly influenced by the polydispersity of the bubble rafts
used.

\begin{figure}
\includegraphics[width=6cm]{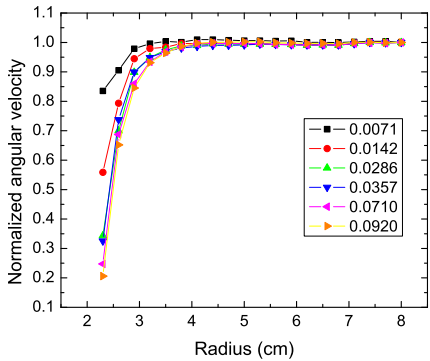}
\caption{(color online)Plot of $v(r)$ for bubbles subject to
Couette shear by driving the outer boundary at a fixed rotation
rate. In this case, the bubbles are open to the atmosphere. The
legend indicates the angular velocity (${\rm s^{-1}}$) of the
outermost layer of bubbles during the applied flow.}
\label{notopprof}
\end{figure}

The average velocity profile of the bubbles in our experiments
have a circular symmetry reflecting the geometry of the shear. The
average radial velocity is zero during steady shear as expected
from the continuity equation in this radially confined system. We
therefore focus on the azimuthal velocity ($v_{\theta}(r)$) as a
function of radial position at different driving rates imposed.
When rotating the outer boundary, a region of ``no-flow''
corresponds to rigid body rotation, $v_{\theta}(r) = \Omega r$,
where $\Omega$ is the angular rotation rate of the outer boundary.
Therefore, to simplify the interpretation of the results, we
present reduced velocities given by $v(r) = v_{\theta}(r)/\Omega
r$, and $v(r) = \rm{constant}$ is a region of ``no flow''. In
Fig.~\ref{notopprof}, we plot $v(r)$ for the case without a top.
It should be noted that instead of using $\Omega$ for the outer
cylinder, we use $\Omega$ of the layer of bubbles adjacent to the
outer cylinder. In the case of no top, these two numbers are
equivalent because there is no slip. However, the presence of a
top introduces slip, and for consistency we found it best to
normalize by the first layer of bubbles.

The measurements of $v(r)$ in the absence of a confining top plate
provide a comparison with previous measurements using a different
Couette apparatus. The main feature of these measurements is the
existence of a shear band at the inner cylinder for all rotation
rates. However, it is worth pointing out that in this regime, the
width of the the shear bands is independent of the rotation rate
of the outer cylinder.

If one defines critical radius for the shear band by the location
at which the flow is indistinguishable from rigid body rotation,
one observes a critical radius on the order of 10 of the smallest
bubble diameter. Alternatively, one can fit the velocity profile
in the shear band region to an exponential, and define a decay
length as in Ref.~\cite{cjwh06}. The results from this fit are
show in Fig.~\ref{decaylengths}. As with the critical radius, the
decay lengths are independent of the angular velocity of the outer
layer of bubbles. Previous studies of bubble rafts without a top
have identified a flow regime for which the critical radius is
independent of the rotation rate of the outer barrier
\cite{gsd06}. This occurs for rates of strain below a critical
value on the order of $0.07\ {\rm s^{-1}}$, and it is referred to
as the discrete regime. Similar behavior has been observed in
three-dimensional foam systems \cite{RBC05}. As the angular
velocity of the outer cylinder is on the order of the rate of
strain, the observation of a rate of strain independent decay
length is consistent with previous studies.

\begin{figure}
\includegraphics[width=6cm]{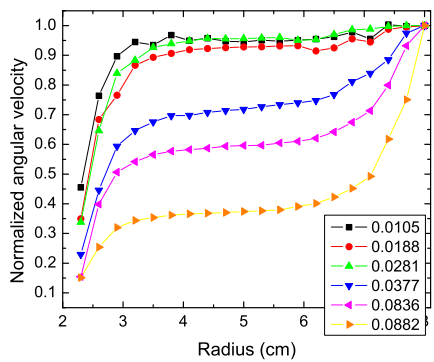}
\caption{(color online) Plot of $v(r)$ for bubbles subject to
Couette shear by driving the outer boundary at a fixed rotation
rate. In this case, the bubbles are confined by a top plate. The
legend indicates the angular velocity (${\rm s^{-1}}$) of the
outermost layer of bubbles during the applied flow.}
\label{topprofiles}
\end{figure}

With the addition of the top plate, the profiles at low rates of
strain are similar to those without a top in that the flow is
localized at the inner cylinder. However, if one considers the
decay length for the shear band, one observes a dependence on the
outer cylinder velocity. (As with the case without a top, an
exponential fit was used to compute the decay length.) As the
velocity increases, the decay length appears to saturate. This
suggests that the dissipation from the top qualitatively changes
the behavior of the system and plays an important role. For
example, this behavior is completely different from what is
observed in the discrete regime, and suggests a continuum model,
such as the one reported on in Ref.~\cite{cjwh06}, may be valid.
Quantitative comparisons to the predictions reported in
Ref.~\cite{cjwh06} are severely limited, in part due to the fact
that the magnitude of the dissipation between the plate and the
bubbles is unknown, and that the model is relatively simplified.
However, it is worth pointing out that the decay length for the
case of rotating the {\it inner} barrier exhibits similar behavior
in the model as observed in our experiments (see Fig.~8 of
Ref.~\cite{cjwh06}). However, this comparison is only indirect, as
we rotate the outer barrier.

As the velocity of the outer barrier is increased, we observe the
behavior predicted in Ref.~\cite{cjwh06}. In addition to flow at
the inner barrier, we observe flow at the outer wall. The region
in between these two zones moves as a rotating rigid body. This
results in a shear band at both boundaries. We were unable to
sample higher shear rates due to limitations of our apparatus, and
have not been able to observe the prediction of a single shear
band at the outer wall. As with the shear bands at the inner
cylinder, we fit the shear bands at the outer wall to an
exponential profile and plot the decay lengths in
Fig.~\ref{decaylengths}. It is interesting to note that the trend
for the outer barrier decay lengths is opposite that of the one at
the inner barrier. Though not reported on in Ref.~\cite{cjwh06},
this is a useful test of the model that can be checked in detail
in the future, and one can make a qualitative argument in the
context of the model proposed in Ref.~\cite{cjwh06}.

As suggested by Clancy et al.~\cite{cjwh06}, the viscous damping
between the top plate and the bubbles is one source of the
formation of shear bands.  In the regime where this dominates,
because the drag increases with increasing velocity, there is a
tendency for the shear bands to move radially outwards with
increasing shear rates. This observation is seen to hold at both
the inner as well as the outer shear bands and is quantified by
the increase in decay length for the inner shear band and
decreasing decay band for the outer shear band with an increase in
the rotation rate. It is also useful to note that while the
viscous drag increases with radial position, the stress increases
with decreasing radial position. The plateau between the shear
bands corresponds to a region where the viscous drag is balanced
by the internal stresses within the bubble raft, resulting in a
constant rotational velocity.

\begin{figure}
\includegraphics[width=6cm]{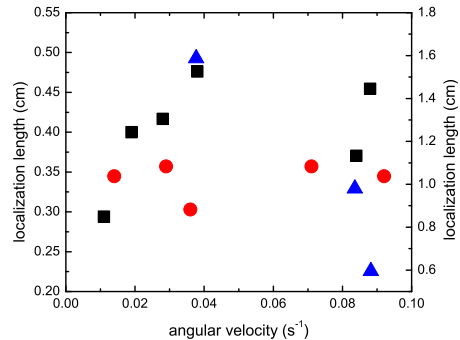}
\caption{(color online) Plot of the decay length for the shear
band as a function of the angular velocity (${\rm s^{-1}}$) of the
outermost layer of bubbles during the applied flow. The results
for the inner shear band are plotted versus the left-hand axis,
with the circles for the system without a top and the squares for
the system with a top. The results for the outer shear band with a
top are plotted versus the right hand axis as triangles.}
\label{decaylengths}
\end{figure}

In summary, the observation of the dual shear bands represents
agreement with the predictions of the theoretical model described
in Ref.~\cite{cjwh06}. For our system, the transition from a
single shear band region at the inner boundary, to a dual shear
band region at the inner and outer boundary occurs at a critical
scaled azimuthal velocity for the outer row of bubbles of $0.03\
{\rm s^{-1}}$. In the context of Ref.~\cite{cjwh06}, this provide
insight into the dissipation due to the top plate. Our previous
work has focused on two cases: bubble rafts with and without a
top. For Couette flow without a top, we have observed two regimes
\cite{gsd06}: discontinuous transitions and a discrete regime. As
discussed, the results for the flow without a top reported here
are consistent with the discrete regime. As such, one does not
expect the model discussed in Ref.~\cite{cjwh06} to be applicable.
Surprisingly, with a top, our current results are consistent with
experiments from a range of groups that have confined bubbles
either between two glass plates or between a fluid and a glass
plate. In all of these cases, exponential profiles for the
velocity are observed \cite{DTM01,wkd06a,KMH08}. The differences
between these studies is the behavior of the width of the shear
band as a function of the applied rotation rate. However, these
differences in rate dependence can be explained in the context of
the continuum model (see for example, Fig.~8 of Ref.~\cite{cjwh06}
and Ref.~\cite{KMH08} which extends the model of \cite{cjwh06}).

Finally, the direct observation of a transition from a single
shear band to a double shear band as a function of increasing
velocity of bubbles for a bubble raft in a Couette geometry
confirms the prediction of the model described in
Ref.~\cite{cjwh06} that such behavior occurs in the presence of
sufficient external drag on a foam system. Experimentally, the
fact that this shear band occurs when a top plate is added
confirms that confining boundaries in quasi-two dimensional foam
systems provide substantial drag that must be understood and
accounted for in theoretical models. Though it remains clear from
other experiments that this mechanism will not explain all cases
of shear-banding, these experiments provide strong evidence that
it is an important mechanism in situations involving external
drag. Some interesting future directions will be to look at the
impact of polydispersity (as in Ref.~\cite{CW08}) and to explore
in more detail the impact of velocity dependent drag forces in the
regime in which discontinuous shear localization exists.

K. Krishan acknowledges financial support by ICAM. M. Dennin
acknowledges the support of University of California, Irvine,
through bridge funding.

%\bibliographystyle{plain}
%\bibliography{kd08}

\end{document}